\documentclass[preprint,amsmath,amssymb,pre,superscriptaddress]{revtex4-1}
\usepackage[utf8]{inputenc}
\usepackage{graphicx}

\usepackage{color}
\usepackage{amsmath}
\usepackage{amssymb}
\usepackage{cases}
\usepackage{indentfirst}
\usepackage{comment} 
\fontfamily{ptm}\selectfont

\begin{document}

\title{A possible four-phase coexistence in a single-component system}
\author{Kenji Akahane}
\affiliation{Institute of Industrial Science, University of Tokyo, 4-6-1 Komaba, Meguro-ku, Tokyo 153-8505, Japan}
\author{John Russo}
\email{russoj@iis.u-tokyo.ac.jp}
\affiliation{Institute of Industrial Science, University of Tokyo, 4-6-1 Komaba, Meguro-ku, Tokyo 153-8505, Japan}
\affiliation{ {School of Mathematics, University of Bristol, Bristol BS8 1TW, United Kingdom} }
\author{Hajime Tanaka}
\email{tanaka@iis.u-tokyo.ac.jp}
\affiliation{Institute of Industrial Science, University of Tokyo, 4-6-1 Komaba, Meguro-ku, Tokyo 153-8505, Japan}

\date{April 26, 2016}

\begin{abstract}
{\bf
For different phases to coexist in equilibrium at constant temperature $T$ and pressure $P$, the condition of equal chemical potential $\mu$ must be satisfied.  This condition dictates that, for a single-component system, the maximum number of phases that can coexist is three. Historically this is known as the Gibbs phase rule, and is one of the oldest and venerable rules of thermodynamics. Here we makes use of the fact that, by varying model parameters, the Gibbs phase rule can be generalized so that four phases can coexist even in single-component systems. To systematically search for the quadruple point, we use a monoatomic system interacting with a Stillinger-Weber potential with variable tetrahedrality. Our study indicates that the quadruple point provides novel flexibility in controlling multiple equilibrium phases and may be realized in systems with tunable interactions, which are nowadays feasible in several soft matter systems (e.g., patchy colloids).}
\end{abstract}
 
\maketitle

%\section*{Introduction}

When different phases are in thermodynamic equilibrium with each other at constant temperature $T$ and pressure $P$, the chemical potentials of the phases must be equal. The number of equality relationships determines the number of degrees of freedom $F$. This leads to the famous Gibbs phase rule \cite{gibbs1961scientific}: $F=C-N+2$, where $C$ is the number of chemically independent constituents of the system, and $N$ is the number of phases. The rule should be valid, provided that the equilibrium between phases is not influenced by external fields and there is no spatial constraint on the phases. The latter condition is known to be violated for coherent solids \cite{johnson1991inapplicability}. This rule tells us that for a pure substance, it is only possible that three phases can exist together in equilibrium ($N = 3$). For a one-component system, there are no degrees of freedom ($F = 0$) when there are three phases ($A$, $B$, and $C$), and the three-phase mixture can only exist at a single temperature 
and pressure, which is known as a triple point. The two equations $\mu_{\rm A}(T, P) = \mu_{\rm B}(T, P) = \mu_{\rm C}(T, P)$ are sufficient to uniquely determine the two thermodynamic variables, $T$ and $P$.
Four-phase coexistence should then be absent, as three chemical potential equations admit no solutions when there are only two independent variables $T$ and $P$.
Mathematically, however, this does not necessarily rule out the possibility that the set of equations may be solved in a special case. Here we seek such a possibility in a systematic manner by tuning the interaction potential, or the Hamiltonian of the system. Extending the dimensionality of the system will allow us to investigate what are the conditions for the existence of a quadruple point. 

The presence of a point where different phases coexist provides an interesting possibility of switching materials properties, 
including electric, magnetic, optical, and mechanical properties, by a weak thermodynamic perturbation such as stressing or heating/cooling.  
The technological importance of a triple point has recently been shown for a popular candidate material for ultrafast optical
and electrical switching applications, vanadium oxide (VO$_2$) \cite{park2013measurement}: It has been revealed that the well-known metal-insulator transition in this material actually takes place exactly at the triple point. Large piezoelectricity near a morphotropic phase boundary is another important example of the importance of multi-phase coexistence \cite{cox2001universal,ishchuk2002investigation,ahart2008origin}. 
In these systems, structural transformations in lattice order are coupled with other orders such as dipole, spin, charge, and orbital, which can be used 
for applications such as electromechanical or magnetoelectronic devices. 
Although the role of multi-phase coexistence in the ease of the transition is not so clear, the minimization of the volume change associated with a phase transition may be realized 
by combined nucleation of two phases with different signs of the volume change upon the transition, which has been reported  
for a transition near a ferroelectric-anitiferroelectric-paraelectric triple point \cite{ishchuk2002investigation}.  
So, the presence of a multiple point may provide a novel kinetic pathway of phase transition, for which the barrier for phase transformation is much lower  
than an ordinary phase transition between two phases.  
Thus, the fundamental understanding of multiple-phase coexistence is not only of scientific interest but also of technological importance.

To study the basics of multi-phase coexistence, we need a model system which shows rich polymorphism. 
In this context, it is well known that water exhibits a rich variety of crystal polymorphs (at least, 16 types of crystals~\cite{russo2014new}). Motivated by this, here we study 
systems interacting with tetrahedral interactions (e.g., covalent bonding and hydrogen bonding).
Tetrahedral interactions are the most important category of directional interactions found in nature,
both in terms of abundance, and in terms of unique physical properties. They are
ubiquitous in terrestrial and biological environments, and fundamental for technological applications. 
The disordered (liquid) phases of tetrahedral materials show unique thermodynamic properties,
the most important being water's anomalies~\cite{debenedetti2003supercooled,russo2014understanding}, like the density
maximum, the isothermal compressibility and specific heat anomaly, etc. 
Ordered phases of tetrahedral materials
are of fundamental importance in industrial application, as they include open crystalline structures, 
like the diamond cubic ($dc$) crystal, or the
quartz crystal, with unique mechanical, optical and electronic properties. 
For example, in Si and Ge, the diamond cubic ($dc$) crystal is a semiconductor, whereas the liquid and body-centred cubic ($BCC$) crystal are metal. Furthermore, $dc$ crystals of mesoscopic particles (like colloids) are also a promising candidate for photonic crystals \cite{maldovan2004diamond}.
It is thus not surprising that tetrahedral interactions are one of the focus of nanotechnology,
with the aim of producing new generation of materials with properties
that can be finely controlled by design~\cite{maldovan2004diamond,Zhang05a,biffi2013phase}.

In order to understand the bulk behaviour of materials with tetrahedral interactions, several coarse-grained
models have been introduced. Among them, probably the most important and successful model is the Stillinger-Weber (SW)
potential~\cite{stillingerweber}, in which tetrahedrality is enforced with the use of three-body force terms. Originally devised as
a potential for silicon, the model has found widespread applicability for several materials, especially
group XIV elements, like germanium and carbon.
The key parameter controlling the tetrahedrality of the model is the ratio between the strength
of three-body interactions over two-body interactions, often referred to as $\lambda$.
As tetrahedrality becomes less strong with increasing atomic number, the basic idea is that group XIV elements, apart from energy and length scale differences, can be modeled by simply varying $\lambda$~\cite{molinero2006tuning}. Even more importantly, the modified SW potential has found general
application as a coarse-grained model for molecular and supramolecular systems. The most important example is water, whose structural properties
have been accurately reproduced with a parametrization of the SW potential (called mW water) with a precision
that is competitive (if not superior) to the best classical molecular models available to date~\cite{molinero2008water,molinero_nature,lu2014coarse}.
  
Despite the importance and widespread applicability of the SW model, our knowledge of its phase diagram
is still lacking. Determining the phase diagram is challenging because
of the three dimensional parameter space (temperature, pressure and $\lambda$), and the fact that all calculations
are multiplied by a large number of crystalline structures with local tetrahedral symmetry that have to be tested for
thermodynamic stability. Previous attempts have considered stable crystalline structures taken from the elements that
the SW potential is ought to describe, as for example silicon.
The first study of the model as a function of $\lambda$ was introduced in Ref.~\cite{molinero2006tuning}
where three crystalline structures were identified at zero pressure $P$: $BCC$,
$\beta$-tin, and $dc$ respectively for low ($\lambda\lesssim 18$), intermediate ($18\lesssim\lambda\lesssim 19$) and high values ($\lambda\gtrsim 19$) of $\lambda$. Interestingly,
the intermediate region showed increased glass-forming ability~\cite{molinero2006tuning}. The $\beta$-tin phase was believed to be the high pressure phase
for SW silicon ($\lambda=21$)~\cite{kaczmarski}. So, according to the current view, starting from a perfectly tetrahedral diamond ($dc$) phase,
the SW system would transform into $\beta$-tin by reducing the amount of global tetrahedrality, either by applying pressure or decreasing the value of $\lambda$.
But this view was recently proven wrong
when a new crystal of SW silicon was found, $sc16$, that replaces $\beta$-tin as the stable phase~\cite{sc16} at high pressures.
$sc16$ is a new crystal with a simple cubic unit cell and 16 atoms in the unit cell.
This calls for a new understanding of the phase behaviour of the SW model, with new
behaviour that should emerge in between
the low $P-\lambda$ region (where $\beta$-tin is thought to be stable) and the high $P-\lambda$ region (where $sc16$ is thought to be stable).

In this Article, we show that indeed the SW model exhibits novel behaviour, which takes the form of a `quadruple point' (QP), where the fluid and
three different crystalline structures ($dc$, $\beta$-tin and $sc16$) have the same chemical potential. 
We also show that the newly found quadruple point is stable against liquid-gas phase separation, by computing the liquid-gas
coexistence line and critical point. At the quadruple point, $\lambda$ takes the value $\lambda_{\rm QP}\sim 20.08$ and the
model describes a one-component system with four-phase coexistence.

\section*{Results}

\subsection*{Modified Stillinger-Weber model}
To compute the phase diagram we run Monte Carlo simulations in the isothermal-isobaric {\it NPT} ensemble~\cite{frenkel2001}.
The SW potential can be written as the sum of a pairwise term $U_2$ and three-body interaction term $U_3$: 
\begin{equation}\label{eq:sw}
U=\sum_{i}\sum_{j>i}U_2({\bf r}_{ij})+\lambda\sum_{i}\sum_{j\neq i}\sum_{k>j}U_3({\bf r}_{ij},{\bf r}_{jk}). \\ 
\end{equation}
Here, $U_2$ models a steep repulsion at short distances and a short-range attraction, while $U_3$ is a directional
repulsive interaction which promotes tetrahedral angles between triplets of particles (for the analytic
expressions of these terms, see Methods).
$\lambda$ is a dimensionless parameter controlling the relative strength between pairwise and three-body term.
Free energy calculations of all relevant crystalline structures are conducted with the Einstein crystal method~\cite{frenkel2001},
and both Gibbs-Duhem integration~\cite{kofke} and Hamiltonian integration~\cite{Vega} are employed to compute coexistence planes and triple lines. Critical points are estimated with grand canonical simulations and histogram reweighting techniques~\cite{reweight}. 
A description of all methods can be found in Methods and from here we use internal units as
explained there.

\subsection*{Phase behaviour of the modified SW model}
We start from a liquid phase and four crystalline phases which are known to be stable for the SW model.
The crystalline phases are body-centered cubic(BCC), $\beta$-tin, diamond cubic(dc), and sc16.
BCC is known as a stable crystalline phase of the SW model at lower $\lambda$~\cite{molinero2008water,molinero2006tuning}.
$\beta$-tin crystal has a body-centered-tetragonal structure with two atoms per cell and is known as a stable crystalline phase for silicon at intermediate pressure~\cite{beta-tin}.
dc is known as a stable crystalline phase for group XIV elements.
sc16 is a crystal which has recently been found to be stable of the SW model at intermediate and high pressure~\cite{sc16}.
The sc16 crystal has a simple cubic unit cell with 16 atoms per cell. The space group of the sc16 crystal is {\it Pa$\overline{3}$}.

\begin{figure*}[h]
 \includegraphics[width=16cm,clip]{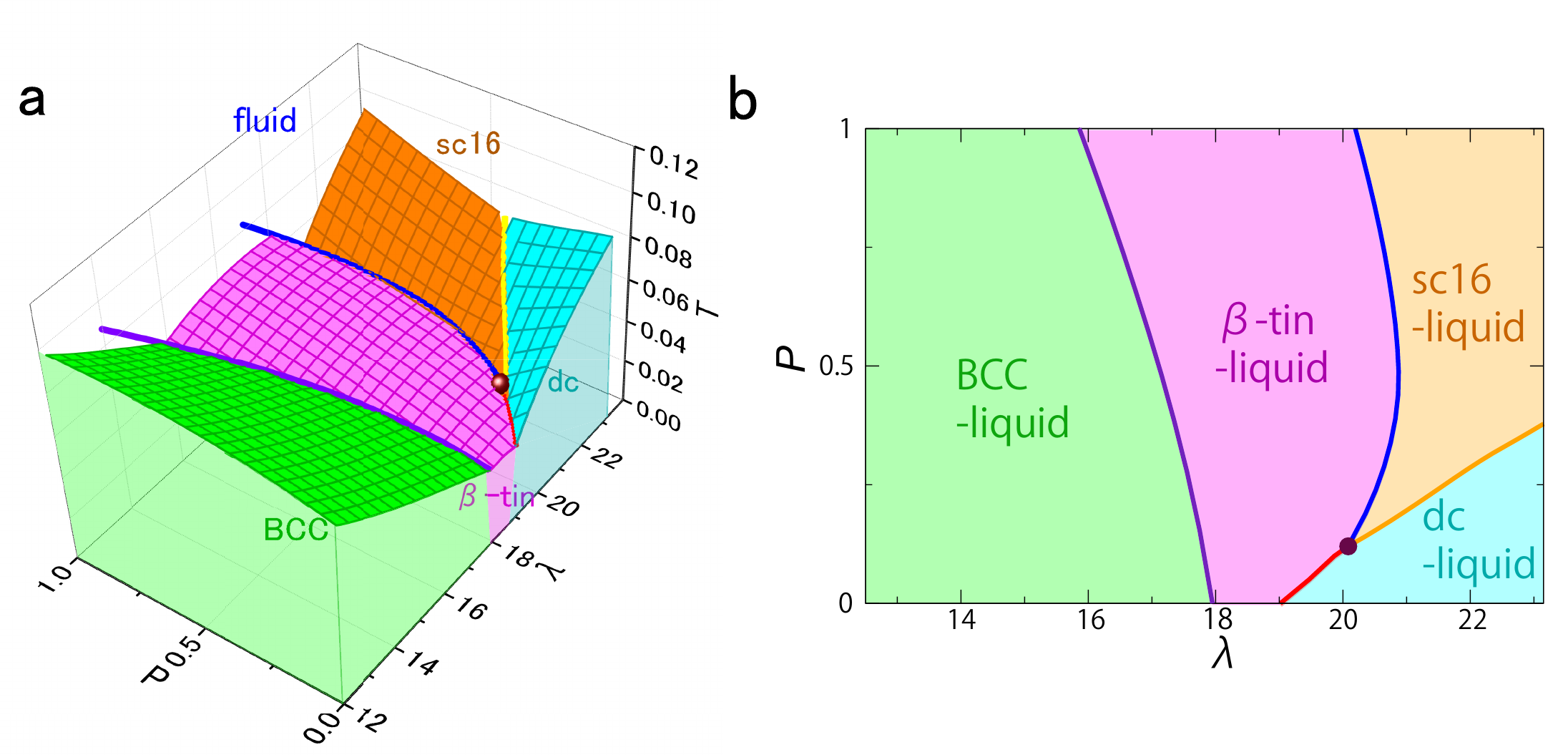}
 \caption{{\bf Phase diagrams of a system interacting with the SW potential.} 
{\bf a,} The $\lambda$-$P$-$T$ phase diagram. 
The green, pink, turquoise and orange surfaces are liquid-BCC, liquid-$\beta$-tin, liquid-dc and liquid-sc16 coexistence surfaces.
The purple, red, yellow and blue lines are liquid-BCC-$\beta$-tin, liquid-$\beta$-tin-dc, liquid-dc-sc16 and liquid-$\beta$-tin-sc16 coexisting lines.
The brown point is a four-phase coexistence point for liquid, $\beta$-tin, dc, and sc16. {\bf b,} The projection of the coexisting regions into $\lambda$-$P$ plane. The green, pink, turquoise and orange regions are the projection of BCC-liquid, $\beta$-tin-liquid, dc-liquid, sc16-liquid surfaces into  $\lambda$-$P$ plane respectively. 
The brown dot is the quadruple point.}
 \label{fig:3dpd}
\end{figure*}

First we show the three-dimensional phase diagram of the SW model, for
$\lambda\in[12.2:23.15]$ and $P\in[0:1]$, in Fig.~\ref{fig:3dpd}a. To aid the visualization, we also plot in Fig.~\ref{fig:3dpd}b a projection of the coexistence
surfaces on the ($P$,$\lambda$) plane. Each surface represents a coexistence surface between
the liquid and the corresponding crystal. Thick lines are triple lines, where two crystalline phases
and the liquid phase coexist. The order of the different crystals is as follows:
BCC (body-centered cubic) at low $\lambda$;
$\beta$-tin at intermediate $\lambda$;
$dc$ and $sc16$, for low and high pressures respectively, at high $\lambda$.
The dot in Fig.~\ref{fig:3dpd} highlights a quadruple point at the intersection
of three triple lines. At the quadruple point $dc$, $\beta$-tin, $sc16$ and the liquid
phase all coexist at the same $T_{\rm QP}$ and $P_{\rm QP}$. The coordinates are approximately:
$\lambda_{\rm QP}=20.08$, $T_{\rm QP}=0.042$ and $P_{\rm QP}=0.120$.
Incidentally, we note that  $\lambda_{\rm QP}$ is very close to the value of $\lambda$ for the SW model of Germanium ($\lambda=20$).
We have checked our results with
direct-coexistence simulations~\cite{vega2008determination}, in which each crystalline phase is placed in contact with the fluid phase
and shown to be at coexistence.

Our results confirm that the $sc16$ is indeed the stable crystalline phase at high ($\lambda$,$P$) and show that
it shares a triple line with the previously known $\beta$-tin phase down to the quadruple point, where the $\beta$-tin
transforms directly into $dc$. We have further confirmed the stability and relevance of the $sc16$ crystalline phase
by direct nucleation events, and showed that the fluid phase directly crystallizes in the $sc16$ phase at
$\lambda=21$, $P=0.5$, and $T=0.0395$. These results are shown in (Supplementary Fig. 1 and 2) and discussed in (Supplementary Note 1).

\begin{figure*}[h]
 \begin{center}
 \includegraphics[width=16cm,clip]{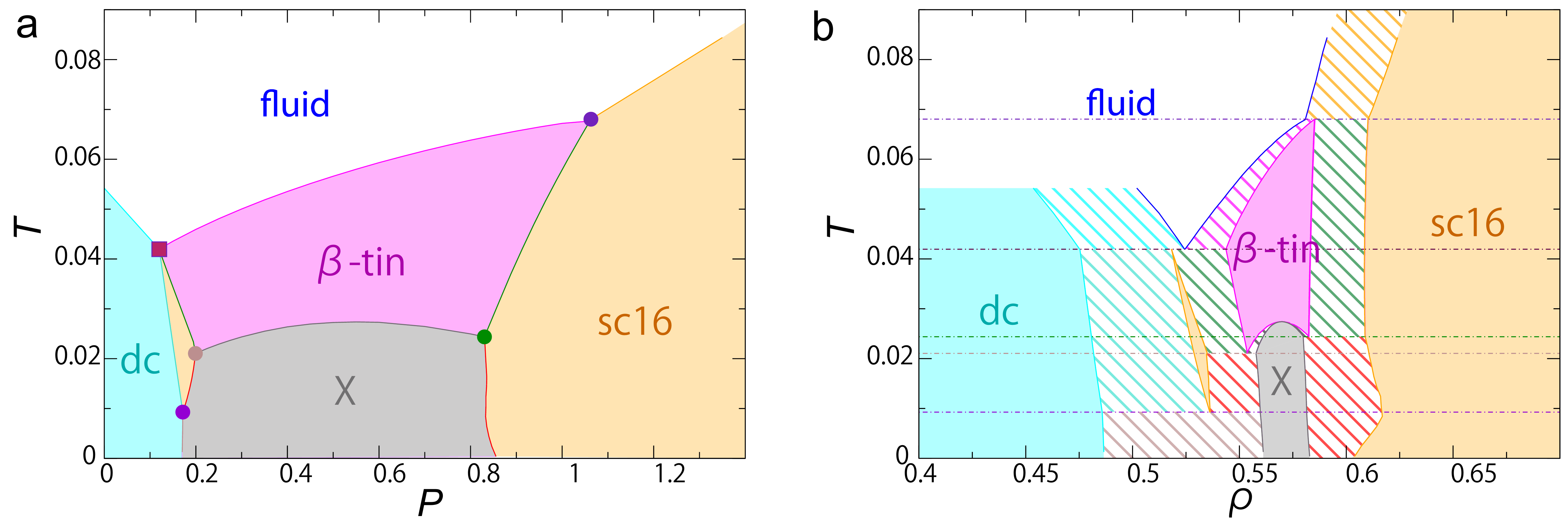}
 \caption{{\bf $P$-$T$ and $\rho$-$T$ phase diagrams of the SW potential at $\lambda\simeq20.08$.} The dc, $\beta$-tin, sc16 and X phases are stable in blue, pink, orange, and gray regions respectively. {\bf a,} $P$-$T$ phase diagram. Here circle points are triple points, and a square point is a quadruple point. {\bf b,} $\rho$-$T$ phase diagram. Regions with diagonal lines are the boundaries of two phases. The horizontal lines indicate temperatures of four triple points and one quadruple point.}
 \label{fig:ptdt20.08}
 \end{center}
\end{figure*}

Unlike $T$ and $P$, $\lambda$ is not a thermodynamic variable but a parameter of the Hamiltonian (Eq.~(\ref{eq:sw})). By choosing
$\lambda=\lambda_{\rm QP}$ we thus have a system with a stable quadruple point in its phase diagram. 
To show this we compute the phase diagrams for $\lambda=\lambda_{\rm QP}$ in the $P$-$T$ and $\rho$-$T$ planes, plotted respectively
in Fig.~\ref{fig:ptdt20.08}a and Fig.~\ref{fig:ptdt20.08}b.
The $dc$ phase is stable at lower $P$ and the $\beta$-tin phase at intermediate $P$.
The stable region of the $sc16$ is instead split into two regions, lower and higher $P$. 
Later, we discuss how the quadruple point emerges when $\lambda\rightarrow\lambda_{\rm QP}$. 
%The quadruple point $T_{\rm QP}$ and $P_{\rm QP}$ originates when, as $\lambda\rightarrow\lambda_{\rm QP}$, the triple points liquid/$dc$/$sc16$ and liquid/$sc16$/$\beta$-tin merge together.

In Fig.~\ref{fig:ptdt20.08}b we show the densities of the different crystalline states.
Diagonal lines represent coexistence regions between two different phases, while horizontal
lines are plotted at the temperatures of the triple points and the quadruple point.
$dc$ is the phase
with lowest density, lower than the fluid's density, as already could be inferred by the slope of the $P,T$ coexistence
line in Fig.~\ref{fig:ptdt20.08}a. The $sc16$ crystal can coexist with the fluid phase at two different densities:
at the quadruple point with a density lower than the liquid's density, and at higher $P$ with a density higher than
the liquid's density.
To summarize our study of the liquid-solid phase diagram, we report all thermodynamic values at triple and quadruple points in Table~\ref{tab:values}.

Figure~\ref{fig:ptdt20.08} shows also the presence of a new phase (denoted as $X$) that we found while computing the 
the coexisting line between $\beta$-tin and $sc16$ at the $\lambda_{\rm QP}$.
The phase spontaneously forms from $\beta$-tin, with which it shares similar densities. As this is a low
$T$ phase which does not coexist with the liquid, and that lies well below the quadruple point, we have not
focused on identifying it. In (Supplementary Fig. 3) and (Supplementary Note 2), we report our preliminary studies on
this new phase, and leave to a future work the task of determining which crystal it represents.

\begin{figure*}[b]
\includegraphics[width=9cm,clip]{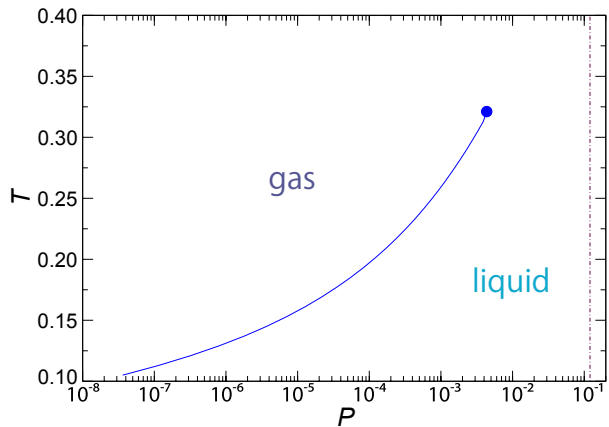}
\caption{{\bf The liquid-gas phase diagram of the SW potential at $\lambda\simeq20.08$.} A circle point is a critical point.
A broken brown line indicates the pressure of the quadruple point.}
 \label{fig:ptgas20.08}
\end{figure*}

\subsection*{Location of the gas-liquid phase transition}
The phase diagrams obtained so far only include solid and liquid phases, so there is the possibility that the quadruple
point is metastable with respect to liquid-gas phase separation. 
To exclude this possibility, we have computed the liquid-gas critical point and coexistence lines for $\lambda=\lambda_{\rm QP}$. 
To obtain the critical point we conducted grand canonical simulations to get the distribution function
of the mixing order parameter $M$ ($M=\rho+su$; $\rho$ is density and  $u$ is internal energy per particle, and $s$ is mixing parameter), and use histogram reweighting to find the state point where this distribution matches the one from the Ising universality class~\cite{reweight,tsypin}. The results for the critical point is $P_{\rm CP}=0.004$, $T_{\rm CP}=0.321$. The gas-liquid phase diagram (the critical point and the coexistence line) is shown in Fig. \ref{fig:ptgas20.08}, 
where it is clear that the liquid-gas critical point is located at pressures
two orders of magnitude lower than $P_{\rm QP}$. Therefore, the quadruple point is indeed a stable thermodynamic point of the model.

\begin{table}[h!]
% \begin{center}
 \begin{tabular}{|c|c|c|c|c|c|c|} \hline
  $T$ & $P$ & $\rho_{dc}$ & $\rho_{\beta-tin}$ & $\rho_{sc16}$ & $\rho_{liquid}$ & $\rho_{X}$ \\ \hline \hline
  0.0093 & 0.171 & 0.486 & & 0.536 & & 0.560 \\ \hline
  0.0211 & 0.199 & & 0.553 & 0.535 & & 0.558 \\ \hline
  0.0244 & 0.831 & & 0.583 & 0.609 & & 0.580 \\ \hline
  0.0420 & 0.120 & 0.475 & 0.544 & 0.518 & 0.525 & \\ \hline
  0.0680 & 1.063 & & 0.585 & 0.610 & 0.581 & \\ \hline
 \end{tabular}
 \caption{{\bf Quadruple and triple points} $T$, $P$ and $\rho$ at triple and quadruple points for the system with $\lambda=20.08$, as shown
 also in Fig.~\ref{fig:ptdt20.08}.}
 \label{tab:values}
%  \end{center}
\end{table}

\subsection*{Emergence of the quadruple point}

\begin{figure*}
 \includegraphics[width=14cm,clip]{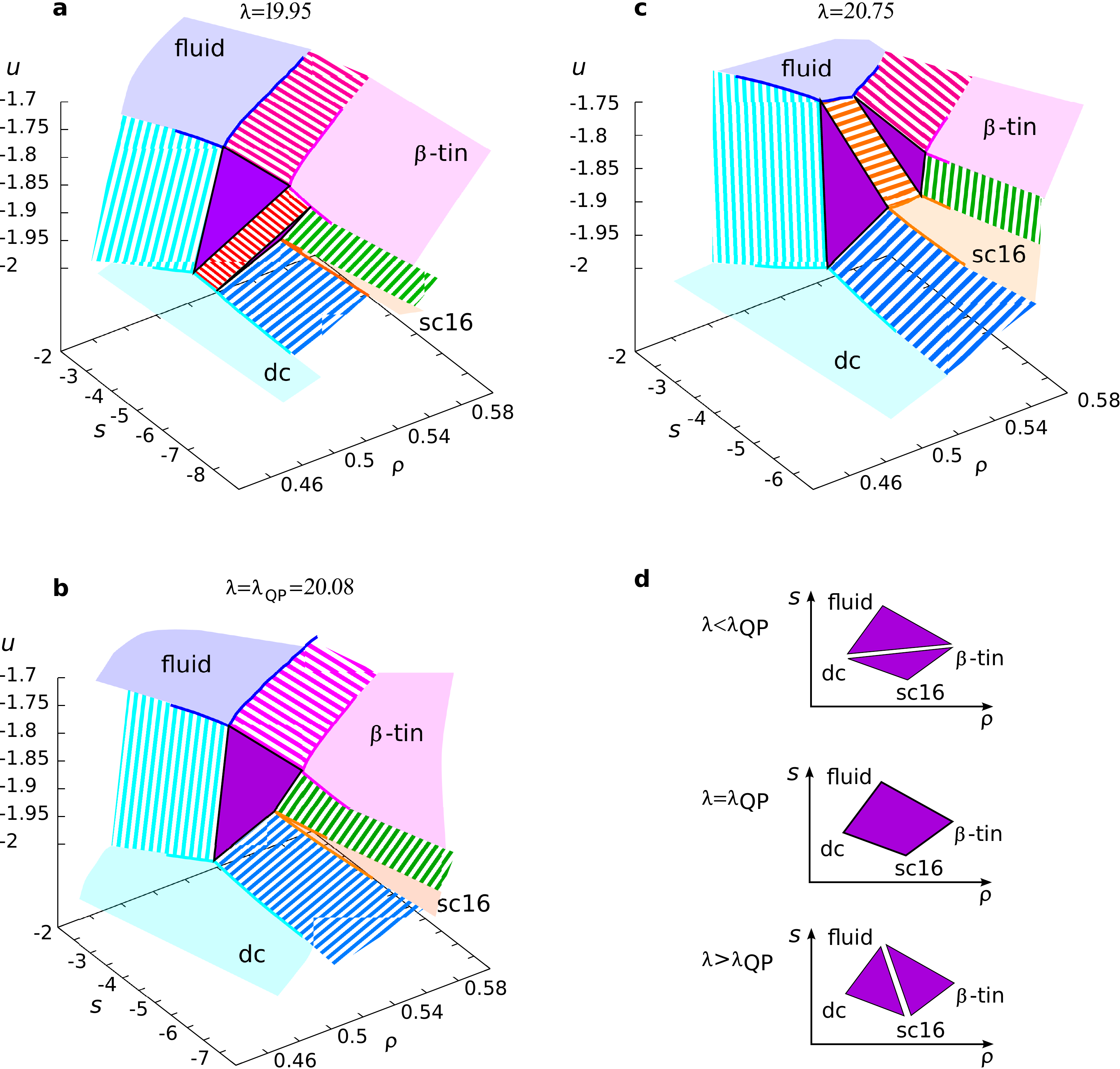}
 \caption{{\bf The emergence of the quadruple point in a single component system.}  {\bf a,b,c} The Gibbs surface
 below ({\bf a}), at ({\bf b}), and above ({\bf c}) the quadruple point.  
 Skew surfaces correspond to pure phases, ruled surfaces to phase coexistence regions, and triangles to triple points. In {\bf b} we show that a quadrangle
 represents four-phase coexistence on the Gibbs surface.
{\bf d,} Schematic figure showing how two triple points merge into a quadruple point and how it splits to form other two triple points 
as a function of $\lambda$.}
 \label{fig:emergence_quadruple}
\end{figure*}%

It is of particular interest to observe how the quadruple point emerges as a function of the parameter $\lambda$ in two dimensions,
where the Hamiltonian is given by Eq.~(\ref{eq:sw}), and $P$ and $T$ are the only intensive thermodynamic variables.
To do this, we consider the Gibbs surface, which expresses all the thermodynamic information on the system as an energy surface $u=u(s,\rho)$
in the entropy $s$ and density $\rho$ plane. In this representation each thermodynamic phase is represented as a surface, whose tangents
are the temperature, $T=(\partial u/\partial s)_\rho$, and pressure, $P=\rho^2(\partial u/\partial \rho)_s/N$.
Triple points are represented as triangles whose vertices lie on the pure phases surfaces.
Slightly below $\lambda_{\rm QP}$, we have two triple points (fluid/dc/$\beta$-tin triple point and  dc/sc16/$\beta$-tin triple point). 
Slightly above $\lambda_{\rm QP}$, on the other hand, we have two different triple points (fluid/dc/sc16 triple point and fluid/$\beta$-tin/sc16 triple point). 
By increasing $\lambda$ continuously from below to above $\lambda_{\rm QP}$, the Gibbs surface shows the following change. 
Below $\lambda_{\rm QP}$ (Fig.~\ref{fig:emergence_quadruple}a), two triangle areas corresponding to the fluid/dc/$\beta$-tin and dc/sc16/$\beta$-tin triple points sandwich 
the dc/$\beta$-tin coexistence surface. With an increase in  $\lambda$, this surface continuously 
becomes narrower, eventually becomes a line, until the two triangles merge forming a quadrangle surface. This quadrangle is made of coplanar
points and is the Gibbs representation of a quadruple point. Figure~\ref{fig:emergence_quadruple}b shows the quadruple point as obtained
from thermodynamic calculations. A further increase in  $\lambda$, leads to the splitting of the quadrangle to two triangles corresponding to the fluid/dc/sc16 triple and fluid/$\beta$-tin/sc16 triple points (Fig.~\ref{fig:emergence_quadruple}c).
The splitting of the quadrangle into two triangles now occurs along the fluid/sc16 coexistence surface. The entire process is schematically
depicted in Fig.~\ref{fig:emergence_quadruple}d.
In two dimensions, the quadruple point is thus formed by the merging of two pairs of triple points located on the same plane in the Gibbs surface.
The degeneracy in the number of degrees of freedom (four phases, but only three equations to determine the volume of each phase)
means that there is no lever rule, and the volume of each phase is not determined by bulk properties alone. The degeneracy is removed if we
consider the system in three dimensions, where $\lambda$ is treated like a thermodynamic parameter (equivalent to an external field).
In this case the Gibbs surface is defined in the three dimensional space of $\rho$, $s$, and $u_3$ (which is the sum of all three body contributions in Eq.~(\ref{eq:sw})).
The quadruple point will be represented as a triangular pyramid, and the volume of each phase at four-phase coexistence is not only determined by the total
volume and entropy, but also by the total three-body energy if we constrain it.

\subsection*{Physical properties associated with the quadruple point}

\begin{figure*}[t]
 \centering
 \includegraphics[width=16cm]{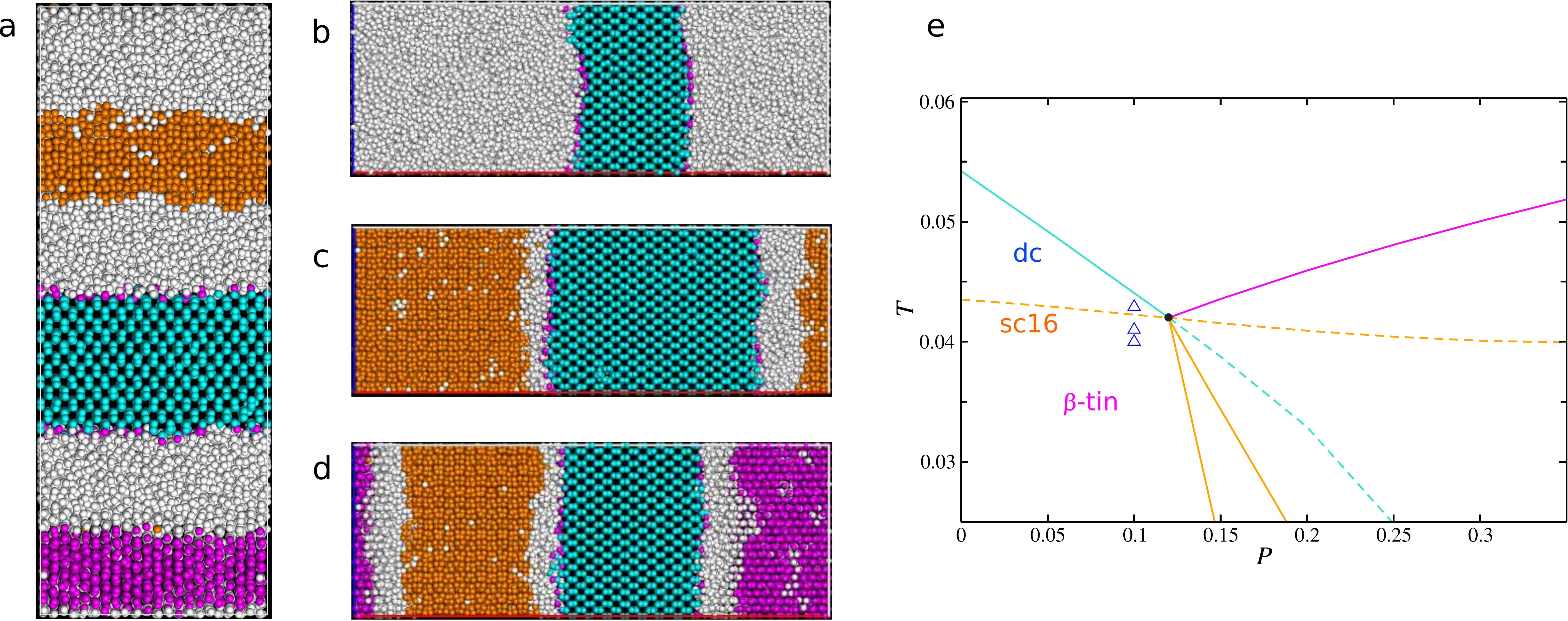}
 % quadruple.pdf: 1585x628 pixel, 72dpi, 55.92x22.15 cm, bb=0 0 1585 628
 \caption{{\bf Direct coexistence near the quadruple point.} {\bf a}, Simulation snapshot of four phases coexisting at
 the quadruple point: liquid (white), dc (cyan), $\beta$-tin (magenta) and sc16 (orange). {\bf b} ,{\bf c}, {\bf d}, Snapshots of configurations
 with one (dc), two (dc and sc16) and three (dc, sc16, $\beta$-tin) crystalline phases respectively, obtained at the following state points near the quadruple
 point, $P=0.1$ and $T=0.0445$ ({\bf b}), $T=0.041$ ({\bf c}), and $T=0.040$ ({\bf d}). {\bf e}, Phase diagram at $\lambda=\lambda_{\rm QP}$, where continuous lines
 are the stable thermodynamic lines of Fig.~\ref{fig:ptdt20.08}{\bf a}, while dashed lines represent their metastable extensions. The black dot is the quadruple point. The triangle symbols indicate the
 state points at which snapshots {\bf b}, {\bf c} and {\bf d} were taken from top to bottom. Note that the liquid layers remain between different types of crystals in panels c and d, despite 
 the liquid is not a stable phase there.}
 \label{fig:quadruple}
\end{figure*}

We next examine some of the unique physical properties associated with the quadruple point.
The simulation shown in Fig.~\ref{fig:quadruple}a was started by interfacing the cubic face of each crystal with a fluid slab,
and equilibrated in the isobaric ensemble where the pressure perpendicular to the interfaces is kept fixed at $P_{\rm QP}$. It is important to ensure that the size of
the interfacial plane is commensurate with an integer number of unit cells for all three crystals at their equilibrium unit cell size. In our simulations
we use $11$, $14$ and $8$ unit cells in each direction of the interfacial plane for $\beta$-tin, dc and sc16 respectively. With this choice we ensure that
all crystals have the correct unit cell size within $\sim 2\%$. The simulations show that the system indeed feels the underlying proximity of a quadruple point, in a way that the free-energy differences between all four phases become negligible, and thus the four bulk phases coexist over practical time scales with free diffusion of all crystal/liquid interfaces (see Fig.~\ref{fig:quadruple}a). 

Next we show that the properties of a quadruple point can be exploited to gain a very fine control over the stability
and number of crystalline phases. Figure~\ref{fig:quadruple}e shows the liquid/solid lines for all crystalline phases, both stable (continuous lines) and metastable
(dashed lines) ones. The quadruple point is the point where all these lines cross. This means that, by choosing thermodynamic conditions arbitrarily close
to the quadruple point, we can stabilize systems with one, two or three crystalline phases. 
This is demonstrated in Fig.~\ref{fig:quadruple}b,c,d where a small
change in temperature allows us to equilibrate systems with varying number of crystalline phases. 
The ability to tune the stability of a varying number 
of crystalline phases with just small changes in thermodynamic parameters is one the most interesting properties of a quadruple point. 
Our simulations also point to the importance of the liquid layer during the solid-solid transition. By placing any couple of crystalline phases directly in contact we always observe the development of several interfacial liquid layers, 
despite the fact that the liquid is not a stable phase in these state points (see Fig. ~\ref{fig:quadruple}c-e). Solid-solid transitions in our systems always proceed in two-steps, where the first step is the melting of the interfacial particles of one solid, and the second step is the nucleation of the second solid phase from this liquid layer. This scenario is similar to what
recently observed in solid-solid phase transitions of colloidal systems~\cite{peng2015two}, and is due to the high interfacial cost of forming a solid-solid interface.

\section*{Discussion}
Quadruple points have not been found before in single-component systems, and the reason
behind their improbability is embodied in the Gibbs phase rule.
Each phase corresponds to a different thermodynamic function (in the isothermal-isobaric ensemble these functions are the chemical potentials of
each phase), and a quadruple point requires the equality of these functions at a single ($T$,$P$) state point. Since the equality
of four different chemical potentials is equivalent to a system of three equations in two variables ($T$ and $P$), the system
should have no solution, and genuine quadruple points should not occur in one-component systems. The way this limitation is
circumvented in our system is by promoting the parameter $\lambda$ to an independent thermodynamic variable, so extending the
dimensionality of space to three dimensions. This allowed us to explore the full ($T$,$P$,$\lambda$) space (Fig.~\ref{fig:3dpd}), where
we were able to find the condition on $\lambda$ for the existence of the quadruple point, where the four phases 
have the same chemical potential.

But, besides the extension of the dimensionality of state-space, a quadruple point
also requires three different stable crystalline phases in a close region of state-space, a condition which is rarely met in
one-component systems. We note that the SW model itself did not meet this requirement until the recent discovery of $sc16$ \cite{sc16}. 

Our work opens for the possibility of realizing a one-component system with a quadruple point in a practical sense, by finding the
conditions at which the strength of the tetrahedral interaction matches the conditions we highlighted in this work. The most likely
candidate for such a system are patchy particles, which are colloidal systems with functionalized patches on their surface~\cite{wang2012colloids,romano2012patterning,rovigatti2014accurate,starr2014crystal,PhysRevLett.115.015701}.
Tuning the angular width of the directional interactions plays a role similar to the parameter $\lambda$ in the SW potential~\cite{romano2010phase,saika2013understanding}, and allows for the tuning necessary to unveil four-phase coexistence.

Here we have determined the full three dimensional phase diagram for the SW model. The relevant
crystalline phases are $dc$, $\beta$-tin, $BCC$, $sc16$ (and the yet unidentified crystalline phase $X$). Apart from $sc16$ and $X$, the remaining phases have been confirmed experimentally for group XIV elements, for which the SW potential is a good coarse-grained model.
The model also displays a quadruple point in the sense that the four phases have the same chemical potential there.
For this model $\lambda=\lambda_{\rm QP}$ we have fully determined the ($P$,$T$) and ($T$,$\rho$) phase diagrams,
and also computed the liquid-gas critical point, showing that the quadruple point is a stable feature of the phase diagram. 
Thanks to the development of technology, a systematic control of the Hamiltonian of a system is now realistic even in experiments:
for examples, ordinary and patchy colloids with more than two types of interactions (see, e.g., \cite{yethiraj2007tunable,maansson2015new}), proteins (see, {\it e.g.}, \cite{bhardwaj2011systematic}), and application of optical and magnetic fields in quantum systems (see, {\it e.g.}, \cite{lukin2000quantum,zaccanti2006control}). 
In such a case, the Gibbs phase rule, which has been considered for two independent thermodynamic variables, should be extended by including 
an additional variable linked to the interaction potential: $F=C-N+M$, where $M$ is a number of independent thermodynamic and Hamiltonian-related variables
~\cite{vega1997plastic}. 
Although we have studied a case of a pure substance ($C=1$ and $M=3$), the above extension of the Gibbs phase rule is not limited to single-component systems. 

Furthermore, by computing the phase diagram as a function of the tetrahedral parameter $\lambda$ our results
can lead to better modelling of atomic systems or enable the design of novel coarse-grained potential 
for new generation materials with directional interactions ({\it e.g.}, patchy particles)~\cite{romano2010phase,romano2012patterning,rovigatti2014accurate,starr2014crystal,PhysRevLett.115.015701}. Our results can also
lead to a better understanding of tetrahedral materials which are arguably the most important class of materials in
nature and technology. For example, our results reinforce the parallelism between pressure and frustration effects,
which is an important principle in the understanding of water mixture and their glass-forming ability (see, {\it e.g.}, \cite{kobayashi2011possible}). 
In the SW model, $\lambda$ controls the degree of deviation from tetrahedrality, and the phase diagram in $\lambda$ has many points in
common with V-shaped phase diagram of real elements in pressure \cite{tanaka2002simple}. For example, the $\lambda$-$T$ phase diagram at ambient pressure, resembles
the the $P$-$T$ phase diagram of tin (Sn), with its succession of grey-tin ($dc$), white-tin ($\beta$-tin) and $BCC$ phases. 

As recently shown in VO$_2$~\cite{park2013measurement} for the ultrafast insulator-metal transition \cite{cavalleri2001femtosecond},
high-order points can be used
to design phase-change materials, where properties change rapidly by applying mechanical stress, heating/cooling, or even tuning the interaction potential (as shown here) by modifying internal degrees of freedom such as spin and electronic states with electromagnetic excitation (see, e.g., \cite{kirilyuk2010ultrafast}).
Thus high-order points open new directions in the control of materials properties. 
We have shown that we can indeed produce any of four different phases by arbitrarily small variations in $T$ or $P$. 
Furthermore, the quadruple point in a single component system should provide great flexibility in controlling multiple phases. 
Here we study phase transitions where the density is the relevant order parameter, but they can be any types of order including dipole, spin, charge, and orbital order (see, e.g. \cite{imada1998metal,dagotto2005complexity}), which are important in functional materials.

\section*{Methods}

\subsection*{Details of the model}

The details of the Stillinger-Weber model are as follows.
The pairwise and three-body interaction terms are given by,
\begin{gather}
U_2(r)=A\epsilon\left[B\left(\frac{\sigma}{r}\right)^p-\left(\frac{\sigma}{r}\right)^{q}\right]\mathrm{exp}\left(\frac{\sigma}{r-a\sigma}\right)  \\
U_3(r_{ij},r_{ik})=\epsilon[cos\theta_{ijk}-cos\theta_{0}]^2\mathrm{exp}\left(\frac{\gamma\sigma}{r_{ij}-a\sigma}\right)\mathrm{exp}\left(\frac{\gamma\sigma}{r_{ik}-a\sigma}\right) 
\end{gather}
where $A=7.049556277$, $B=0.6022245584$, $p=4$, $q=0$, $cos\theta_{0}=-1/3$, $\gamma=1.2$, $a=1.8$.
The parameter $\epsilon$ sets the energy scale and $\sigma$ the length scale.
They correspond to the depth of the two-body interaction potential and the particle diameter respectively, and determined by materials for which the model is used.
We use internal units where $\epsilon$ and $\sigma$ are the units of energy and length respectively.
Therefore, $\lambda$ is only parameter which differentiates the models.

\subsection*{Simulation methods}

We compute internal energies and densities at each equilibrium state by performing Monte Carlo simulations.
The size and shape of the simulation box can fluctuate so as to allow crystalline phases to change their structures \cite{boxshape1,boxshape2}.
A volume-change attempt occurs on every $N$ translation attempts.
The number of particles in the box is 1024, which is large enough so that finite-size effects are negligible.

We obtained the phase diagram by extending coexisting lines from phase diagrams at zero pressure and at $\lambda=23.15$ \cite{sc16}.
We extend two-phase coexisting lines in two directions, along pressure axis and along $\lambda$ axis.
We perform Gibbs-Duhem integration and Hamiltonian Gibbs-Duhem integration~\cite{Vega, kofke}
in order to obtain coexisting lines along pressure axis and along $\lambda$ axis respectively.

When we compute triple lines, we use the following relationships.
For three phases 1,2,3 at coexistence,
\begin{numcases}
{}
v_1dp-s_1dT+\left(\frac{\partial g_1}{\partial \lambda}\right)d\lambda=v_2dp-s_2dT+\left(\frac{\partial g_2}{\partial \lambda}\right)d\lambda   \\
v_1dp-s_1dT+\left(\frac{\partial g_1}{\partial \lambda}\right)d\lambda=v_3dp-s_3dT+\left(\frac{\partial g_3}{\partial \lambda}\right)d\lambda 
\end{numcases}
$v$, $s$ and $g$ are the volume, entropy and Gibbs free energy per particle.
By solving this set of equations, we get
\begin{numcases}
{}
\frac{dT}{d\lambda}=\frac{(v_1-v_3)(w_1-w_2)-(v_1-v_2)(w_1-w_3)}{(v_1-v_3)(s_1-s_2)-(v_1-v_2)(s_1-s_3)} \\
\frac{dp}{d\lambda}=\frac{(s_1-s_3)(w_1-w_2)-(s_1-s_2)(w_1-w_3)}{(s_1-s_2)(v_1-v_3)-(s_1-s_3)(v_1-v_2)} 
\end{numcases}
Here, we write
\begin{equation}
w=\left(\frac{\partial g}{\partial \lambda}\right). \nonumber
\end{equation}
When the internal energy per particle $u$ can be written as $u=u_{\rm a}+\lambda u_{\rm b}$, we can use~\cite{frenkel2001,Vega}
\begin{align}
w&=\left(\frac{\partial g}{\partial \lambda}\right)=\langle u_{\rm b}(\lambda) \rangle_{N,p,T,\lambda}. 
\end{align}
Here $\langle ... \rangle_{N,p,T,\lambda}$ means ensemble average with constant $N,p,T,\lambda$, which can be determined within an {\it NPT} simulation.
By integrating these equations, we extend triple points.

In order to obtain critical points, we compute distribution functions of densities and energies with using histogram reweighting method~\cite{reweight}
and fit them according to the universal Ising universality class~\cite{tsypin}.

\section*{References}

%\bibliographystyle{naturemag2}
%\bibliography{biblio}

\section*{Acknowledgments}
The authors thank F. Romano for helpful discussions. This study was partly supported by Grants-in-Aid for Scientific Research (S) (Grand No. 21224011) and Specially Promoted Research (Grand No. 25000002) from the Japan Society for the Promotion of Science (JSPS). 

\clearpage

\setcounter{figure}{0}
\centerline{\bf \large Supplementary Information}

\renewcommand{\figurename}{{\bf Supplementary Figure}} 
\renewcommand{\thefigure}{\arabic{figure}} 
\begin{figure}[h!]
\begin{center}
 \includegraphics[width=12.cm,clip]{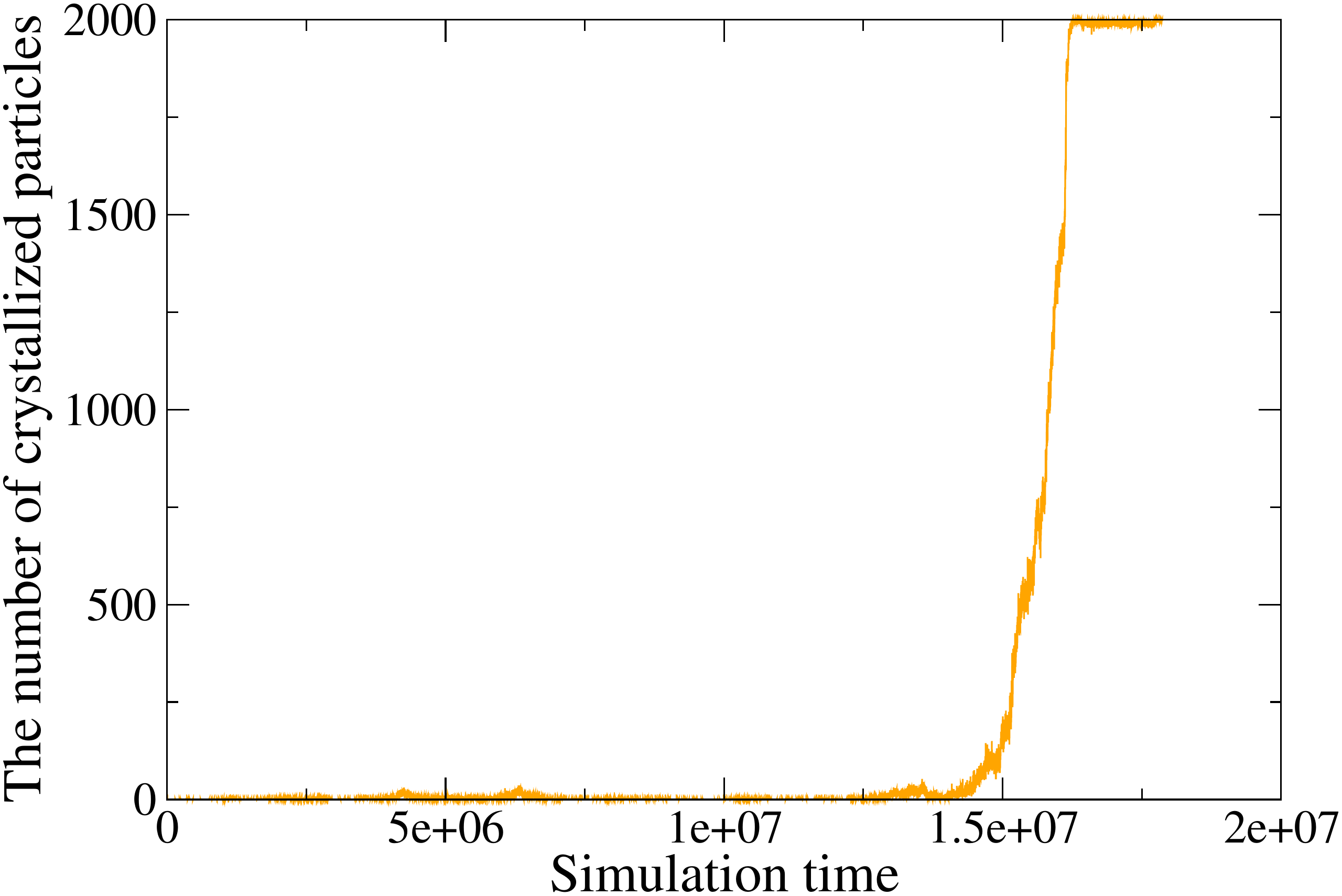}
\end{center}
 \caption{{\bf The number of particles crystallized to sc16 as a function of time.} The total number of particles in the simulation box is 2000. }
 \label{fig:op}
\end{figure}%

\begin{figure}[h!]
\begin{center}
\includegraphics[width=16cm,clip]{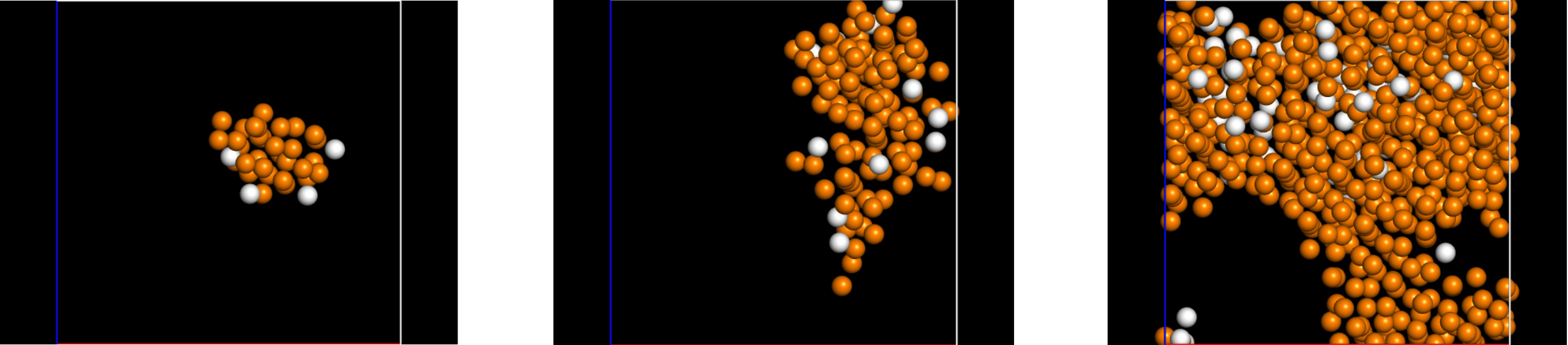}
\end{center}
\caption{{\bf The configurations of particles crystallized to sc16 during crystallization.} 
Orange particles have the symmetry of sc16, whereas white particles represent defects.}
\label{fig:trajectory}
\end{figure}%
%\end{spacing}
%\clearpage
\begin{figure}[h!]
 \centering
\includegraphics[width=10.0cm]{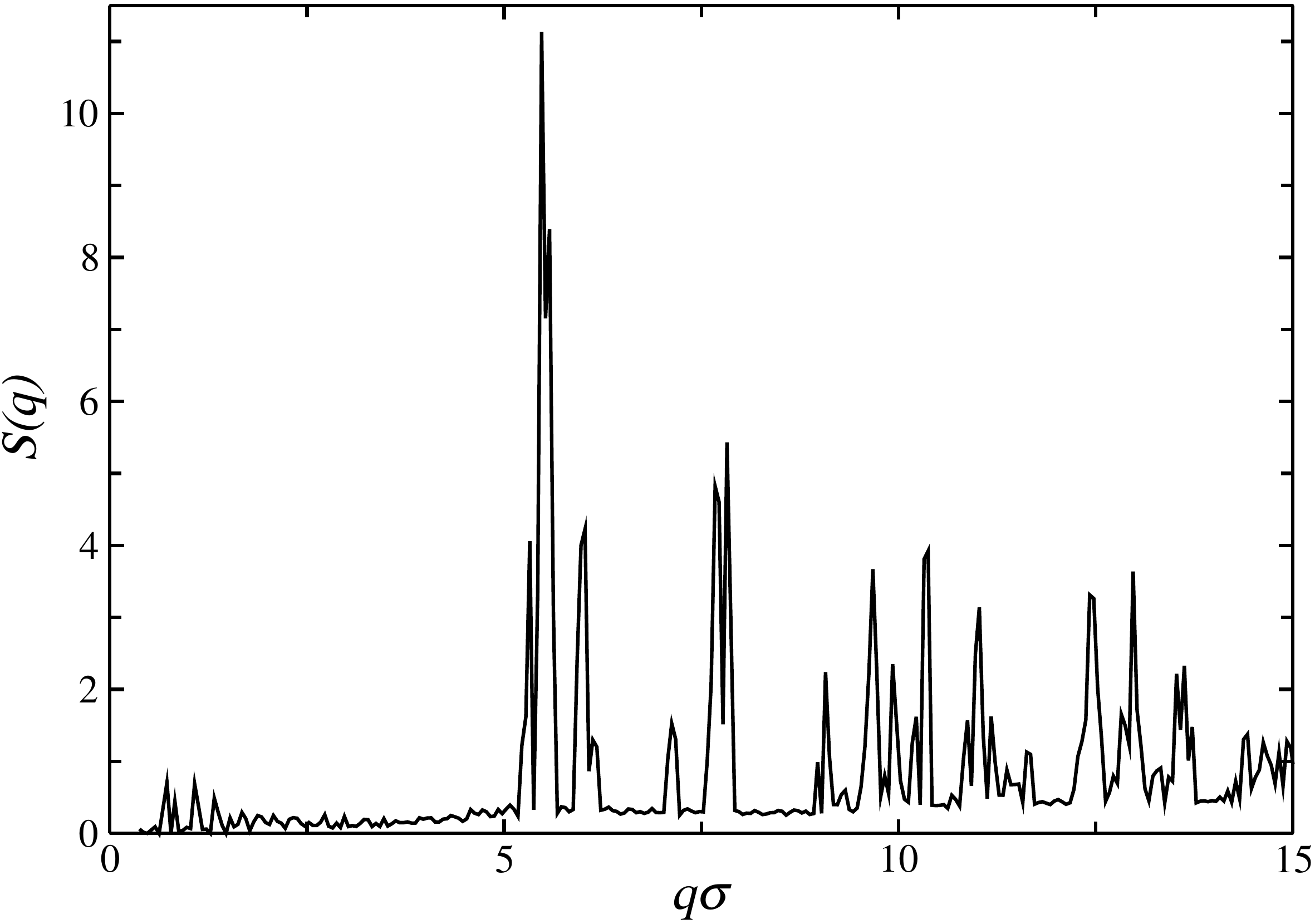}
 % sq.eps: 0x0 pixel, 300dpi, 0.00x0.00 cm, bb=(atend)
 \caption{{\bf Structure factor of the X phase.} Static structure factor of the X phase.}
 \label{fig:sq}
\end{figure}
%\clearpage
\noindent
\vspace{1cm}

\section*{Supplementary Note 1: Direct nucleation of the sc16 phase}

In order to confirm that fluid phase directly crystallizes in the sc16 phase, we perform Monte Carlo simulations in isothermal-isobaric ensembles
at $\lambda=21$, $P=0.5$, and $T=0.0395$.
We use a liquid phase as initial configurations and run 50 independent trajectories, which all crystallize.
Supplementary Figure~\ref{fig:op} shows the number of particles crystallized to sc16 for one of these trajectories, and Supplementary Figure~\ref{fig:trajectory} shows 
the configurations of particles crystallized to sc16.

\section*{Supplementary Note 2: The X phase}

As discussed in the main manuscript, we have found a new stable crystal into which the $\beta$-tin phase transforms at
low-temperatures. We have confirmed that this phase is not one of the other phases considered in this work. In Supplementary Figure~\ref{fig:sq}
we plot the structure factor obtained from equilibrated configurations of the crystal. We leave for future work the classification of this crystalline phase.

\end{document}